\documentclass[english, a4paper, 12pt]{article}
\usepackage{graphicx,psfrag}
\usepackage[T1]{fontenc}
\usepackage[margin=1in]{geometry}
\usepackage{hyperref}
\usepackage[round,authoryear]{natbib}
\usepackage{bibunits}
\usepackage{graphicx, color}
\usepackage{epsfig}

\title{The ``Size Premium'' in Equity Markets:\\ Where is the Risk?}
\author{Stefano Ciliberti, Emmanuel S\'eri\'e, Guillaume Simon,\\ Yves Lemp\'eri\`ere \& Jean-Philippe Bouchaud
\\Capital Fund Management, \\ 
23 rue de l'Universit\'e, 75007 Paris, France}
\date{\today}

\begin{document}

\maketitle
\begin{abstract}
We find that when measured in terms of dollar-turnover, and once $\beta$-neutralised and Low-Vol neutralised, the Size Effect is alive and well. With a long term t-stat of $5.1$, the ``Cold-Minus-Hot'' (CMH) anomaly is certainly not less significant than other well-known factors such as Value or Quality. As compared to market-cap based SMB, CMH portfolios are much less anti-correlated to the Low-Vol anomaly. In contrast with standard risk premia, size-based portfolios are found to be virtually unskewed. In fact, the extreme risk of these portfolios is dominated by the {\it large cap} leg; small caps actually have a positive (rather than negative) skewness. The only argument that favours a risk premium interpretation at the individual stock level is that the extreme drawdowns are more frequent for small cap/turnover stocks, even after accounting for volatility. This idiosyncratic risk is however clearly diversifiable.
\end{abstract}

\section{Introduction}

\begin{figure}[!htb]
\centering
\includegraphics[width=10cm]{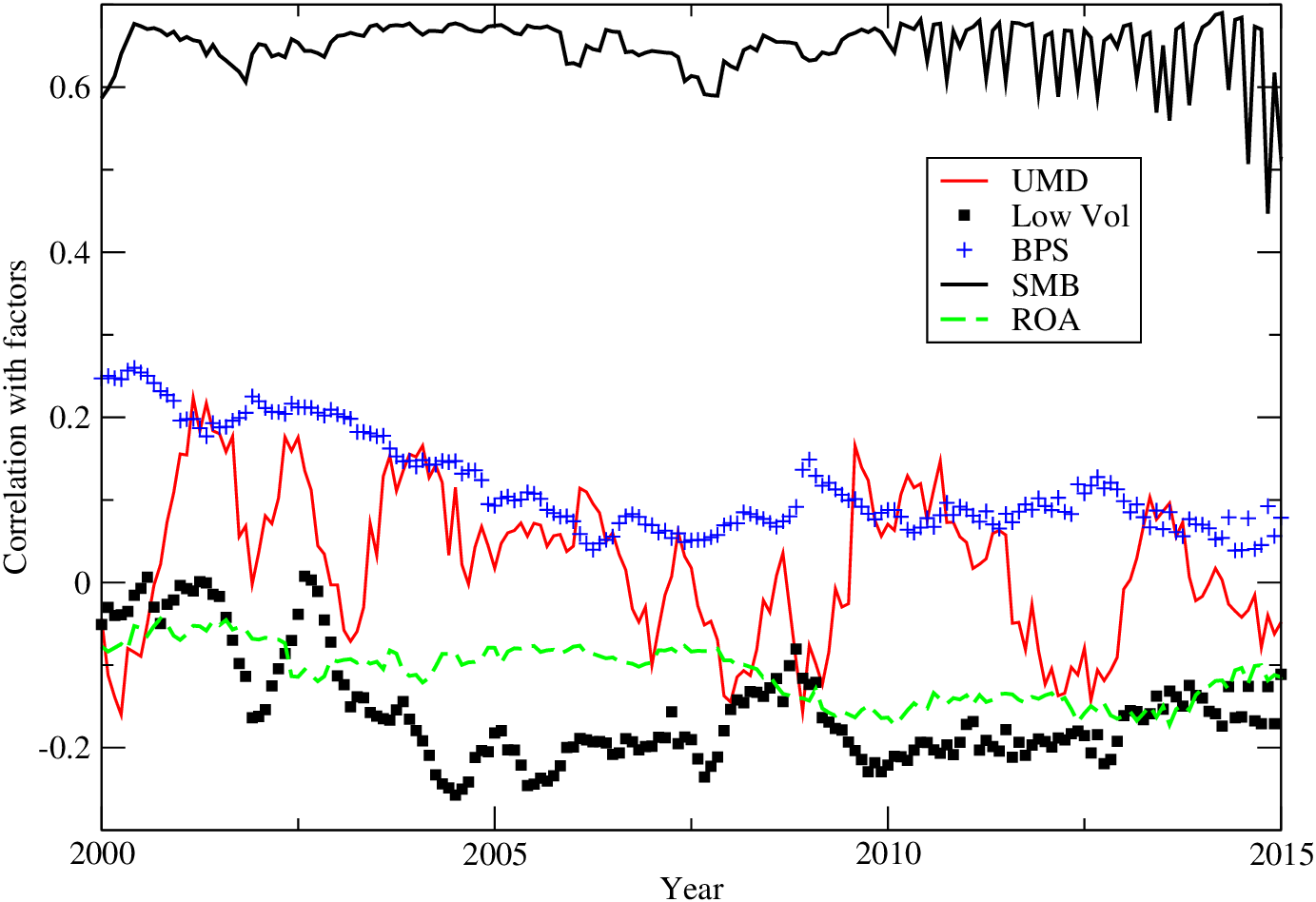}
\caption{Correlation of the average fund position with the major equity factors (UMD, LowVol, BPS, SMB, ROA) as a function of time. The high correlation with SMB shows that the typical US mutual fund has a strong small-cap bias in its equity selection process. The average exposure to other equity factors is much smaller in absolute terms, and fluctuates more with time. Source: FactSet data on Mutual Fund positions (US). UMD: Up-Minus-Down, aka ``Momentum''; BPS: Book-per-Share; ROA: Return-over-Assets.}
\label{PopularFactor}
\end{figure}

One of the best known -- and perhaps most controversial -- effects in the market folklore is the so-called ``size premium'', which states that small-cap stocks are on average undervalued and outperform large-caps. But while still highly popular among equity managers (see Figure \ref{PopularFactor} for an illustration), strong doubts about the very existence of a size premium have been expressed in a number of scientific publications. The first discussion of the ``Small (cap) Minus Big (cap)'' (aka SMB) effect is often attributed to study by Banz in 1981 \cite{Banz1981}, which was followed by a host of publications (e.g., \cite{Roll1981}, \cite{Reinganum1981}, \cite{Christie1981}, \cite{Schwert1983}, \cite{Blume1983}, \cite{ChanHsieh1985}, \cite{Huberman1987}, \cite{ChanChen1988}), culminating with the famous Fama \& French paper \cite{FamaFrench1993} where the size premium was minted into a fundamental ``risk factor'', along side with the Market factor and the Value factor. 

However, the more recent literature has not been kind to SMB, see e.g. \cite{vanDijk2011}. The effect is often dismissed as a ``myth'' or a statistical fluke, which at best existed before Banz's paper, but has disappeared now it has become common lore. But since small cap stocks should indeed be riskier (in a sense made precise below), where has the risk premium gone? A recent paper by Asness and associates \cite{Asness2015} argues that ``\emph{size matters, if you control your junk}''; i.e., that small cap firms overperform large cap firms when one restricts to high quality firms. This suggests in effect that the size premium can be masked by ``junk''. 

Here we want to advocate another, complementary picture. We argue that market capitalisation is not an optimal indicator of an otherwise genuine ``size'' effect. Indeed, the dependence of a stock ``beta'' on market capitalisation is \emph{non-monotonic}, which induces spurious biases in a (market neutral) portfolio construction. The resulting SMB portfolios have a strong short Low-Vol exposure. We propose instead the average daily volume (ADV) of transaction (in dollars) as an alternative indicator of size, and show that the above mentioned biases are substantially reduced. The choice of ADV is further motivated by two independent arguments, often put forth in the literature. One is that the size effect might in fact be a liquidity risk premium, i.e., ``Cold'' stocks, that are more difficult to liquidate, are traded at a discount. The other is that heavily traded, ``Hot'' stocks are scrutinized by a larger number of market participants, therefore reducing pricing errors. 

Although these arguments sound reasonable, we will show below that they fail to capture the mechanism underlying the profitability of Cold-Minus-Hot (CMH) portfolios. Perhaps surprisingly, standard skewness measures do not conform to the idea of that the size effect is a risk premium. In fact, the single name skewness of small cap/small ADV stocks is \emph{positive}, and declines to zero as the market cap/ADV increases. At the portfolio level, small cap/ADV stocks do \emph{not} contribute to skewness either. In fact, the SMB portfolio is only weakly negatively skewed, whereas the CMH portfolio is not skewed at all; furthermore, large gains/losses at the portfolio level mostly come from the short leg (corresponding to large cap/ADV stocks). All these results suggest that ``prudence'' (i.e. aversion for negative skewness and appetite for positive skewness) should favour small cap/ADV stocks, in contradiction with the idea that SMB or CMH are risk premia strategies. 

Interestingly, however, higher moments of the return distribution, such as the kurtosis or the downside tail probability, show a clear decreasing pattern as a function of market cap or ADV. In other words, extreme in both directions are more common for small stocks, even after factoring out volatility. Even if, quite unexpectedly, large upside events are more likely than large downside events for small stocks, ``safety first'' considerations might be enough to deter market participants from investing in these stocks. This scenario would allow one to think of the size anomaly as a risk premium -- albeit a rather non conventional one.

\section{Beta-neutrality Distortions \& Low-Vol Biases}

\begin{figure}[!htb]
\centering
\includegraphics[width=10cm]{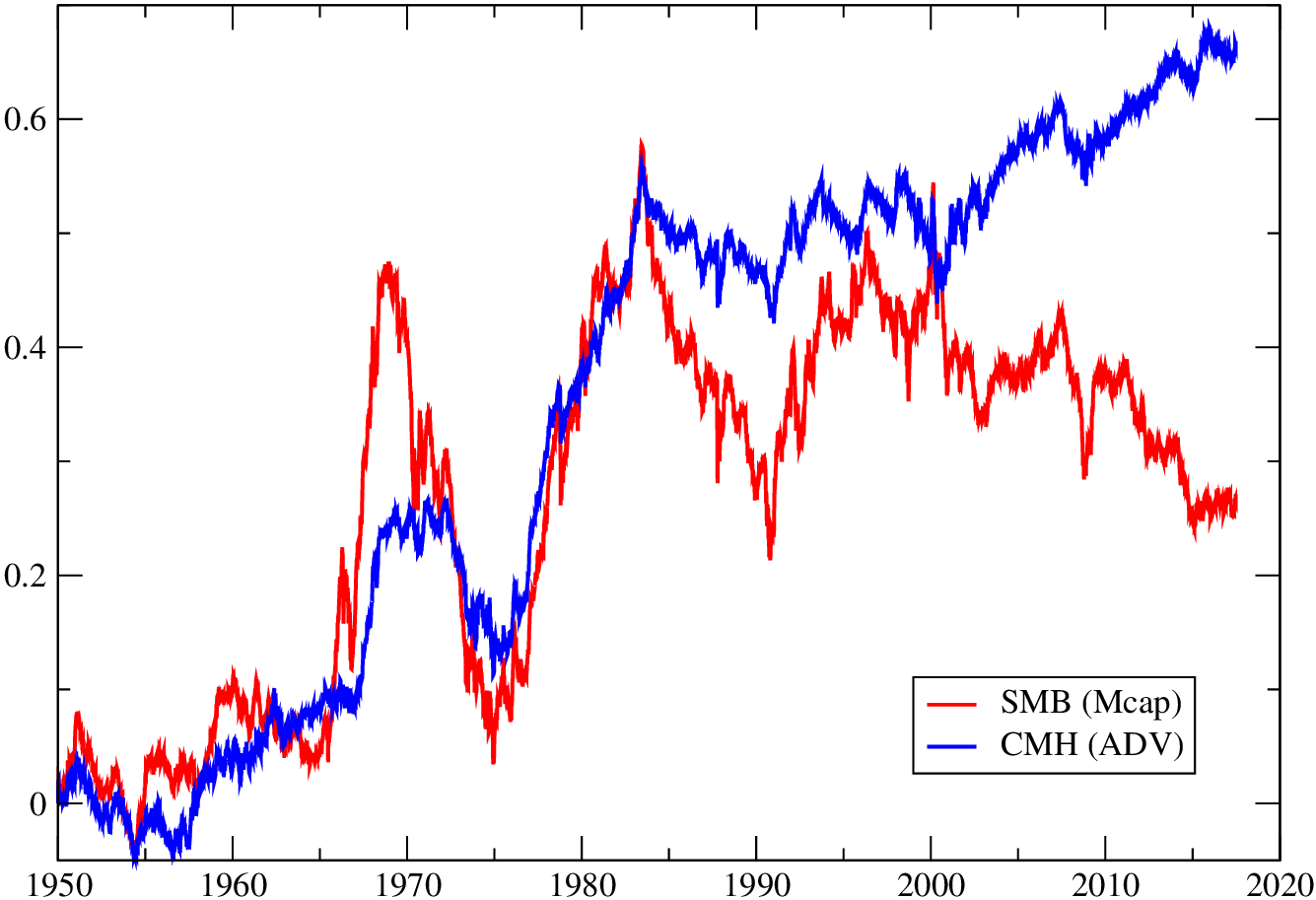}
\caption{Worldwide Simulation of SMB (based on Market Cap) and CMH (based on ADV). The pool includes the following countries (with their entry date in the pool in parenthesis): Australia (1996-01), US--CRSP (1950-01), US--non CRSP (1990-01), Europe (1996-01), Japan (1993-01), Korea (2002-04), Taiwan (1999-01), Hong-Kong (2002-01), Brazil (2001-01), Canada (2001-04).}
\label{WWSimulation}
\end{figure}

The standard Fama-French SMB factor suffers from a strong market bias, which pollutes the performance of SMB portfolios. In our work, we consider a continuous SMB signal in $[-1,1]$ based on the rank of the market capitalisation, rebalanced every day, which has an average $\beta$ of +7 \%.\footnote{Here and throughout this paper, the $\beta$ is computed as the correlation with the market cap weighted index.} Once $\beta$-neutralised, we find that the SMB effect, diversified across a large international pool, is not significant: its t-stat is $0.9$ since 1950 (see Figure \ref{WWSimulation}). This very low t-stat value is at the origin of the current debate about the very existence of a size effect.

\begin{figure}[!htb]
\centering
\includegraphics[width=8cm]{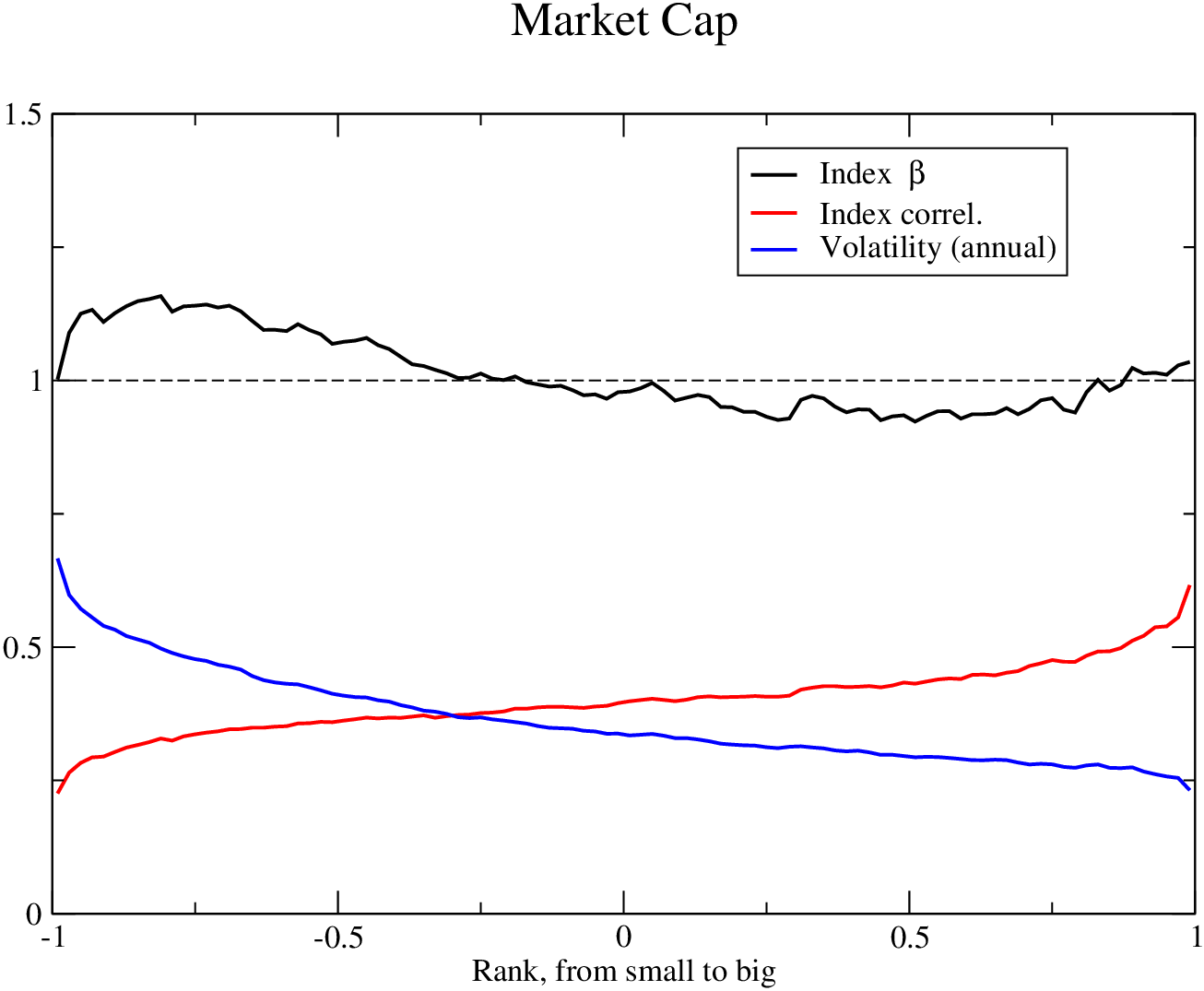}\hskip 0.5cm\includegraphics[width=8cm]{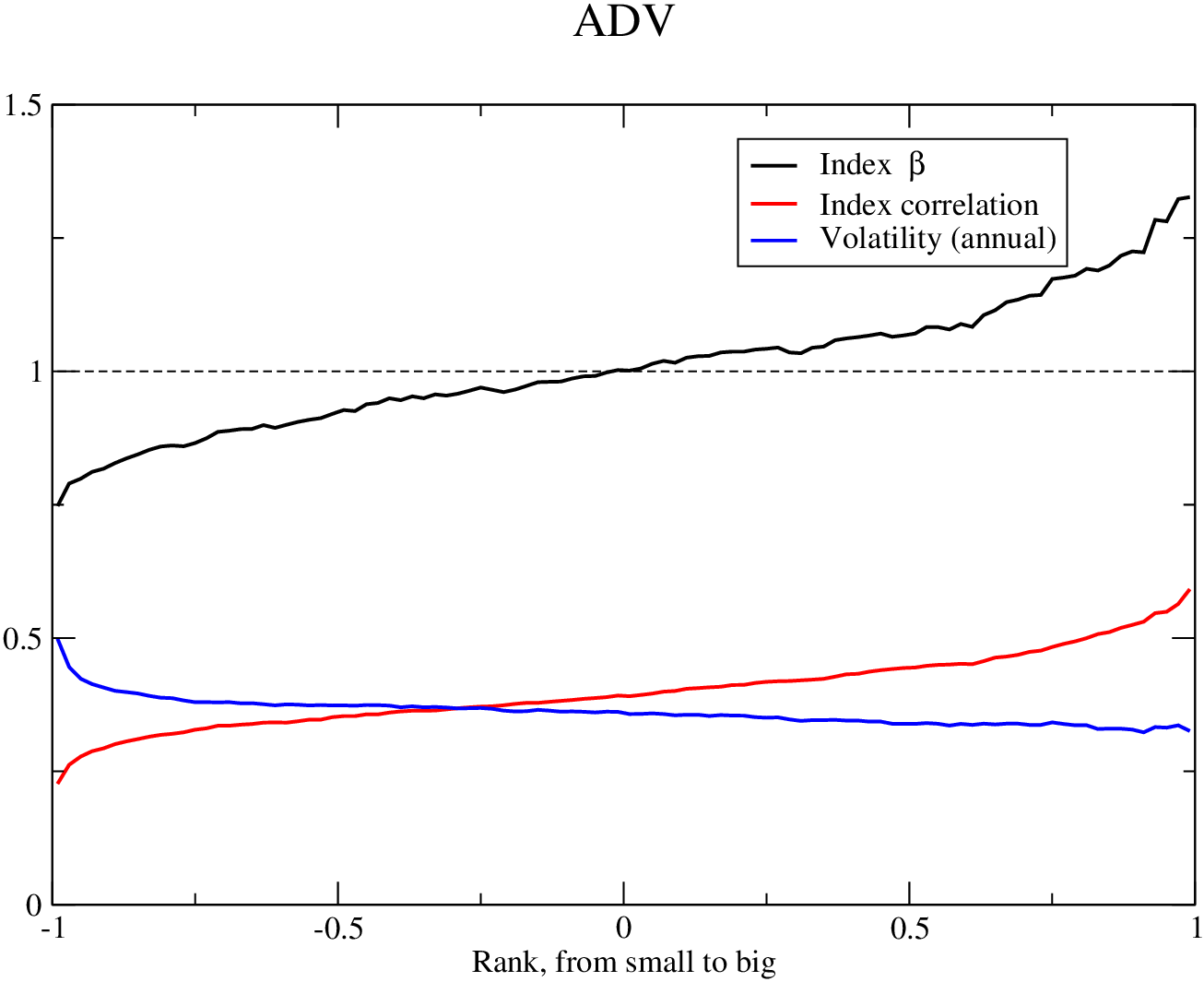}
\caption{Individual volatility, beta and correlation (wrt to the market cap weighted index) as a function of (top) ranked market capitalization and (bottom) ADV. Note that the dependence of beta on market cap is non-monotonic, but becomes much better behaved as a function of ADV.}
\label{BetaExplosurePlot}
\end{figure}

As revealed by Figure \ref{BetaExplosurePlot}, the $\beta$-bias is in fact quite non-trivial. The dependence of the $\beta$ of individual stocks on market capitalisation is, perhaps surprisingly, non-monotonic: very small caps and large caps have a $\beta$ smaller than 1, whereas medium caps have a $\beta$ larger than 1. This non-monotonic behaviour comes from the competition of two opposite effects: volatility tends to decrease with capitalisation, whereas correlation with the index increases with capitalisation (see Figure \ref{BetaExplosurePlot} (top)). The $\beta$ is the product of these two quantities, and therefore exhibits a hump-shaped pattern. Any $\beta$-neutralisation will therefore lead to a market cap distortion in the final portfolio, and potentially obfuscate the size effect. 

Another obvious consequence of the volatility vs. cap pattern is that SMB portfolios are anti-correlated ($\approx -80 \%$) with Low-Vol portfolios (i.e., portfolios that invest in low-volatility stocks and short high-volatility stocks). Since the Low-Vol effect is statistically significant, such a negative exposure must degrade the performance of SMB. Once both the $\beta$ and the volatility biases are removed, the t-stat of a diversified SMB portfolio indeed rises to a nifty value of $5.0$.

\section{Correcting Biases: the CMH Portfolio}

There are obviously many ways to assess the ``size'' of a company, market capitalisation being only one of them. An enticing candidate is the average daily volume (ADV), or more precisely the total amount of daily traded dollars in a given stock. This is both a measure of the liquidity of a stock (i.e. the possibility of rapidly unwinding a large position with relatively small impact costs) and of the overall interest of the market for the stock. Although quite natural, this quantity is only rarely considered as a relevant indicator (but see \cite{BrennanSubra1996}). In fact, the closest to ADV is Amihud's liquidity indicator LIQ, defined as ADV divided by daily volatility \cite{Amihud_02} but as we will see later this mechanically re-introduces a negative exposure to Low-Vol. 

Interestingly, when plotted as a function of the ADV instead of the market capitalisation, the evolution of the $\beta$ of individual stocks is now found to be monotonic and increasing, see Figure \ref{BetaExplosurePlot} (right). When used to construct $\beta$-neutralised ``CMH'' portfolios (constructed as the SMB portfolios, with a signal in $[-1,1]$ based on the rank of the ADV), we find a very significant improvement of the t-stat that jumps from $0.9$ to $3.2$. The Sharpe ratio is only negative in Canada, with a t-stat of $-2.0$. Notably, both the short leg (large ADVs) and the long leg (small ADVs) are profitable after a separate $\beta$-neutralisation of the two legs, whereas the ($\beta$-neutralised) large cap leg of SMB has in fact a \emph{negative} performance with a t-stat of $-2.3$.  This shows that the ADV ranking contains some information, even restricted to the large- and small-ADV sub-pools. 

Note furthermore that the (anti-)correlation between volatility and ADV is much weaker (US: - 14 \%, WW: - 14 \%) than the correlation between market cap and volatility (US: - 55 \%, WW: - 47 \%), see Figure \ref{BetaExplosurePlot}. Once the Low-Vol component of our CMH portfolio is subtracted, the t-stat is increased to $5.1$.\footnote{Because the correlation with Low-Vol is smaller, the relative increase of t-stat is weaker for CMH than for SMB.} Therefore, we conclude that there is indeed a genuine, highly significant size effect when measured using ADV, and once $\beta$ and Low-Vol exposure are removed. This, in our view, is an important conclusion, which complements the findings of Asness et al. \cite{Asness2015}. We in fact  confirm the negative ($- 16\%$) correlation between the performance of a ``Quality'' portfolio (based on Net Income over Total Assets) and our CMH portfolio. In combination, CMH and Quality lead to a performance with an engaging world-wide t-stat of $5.7$ since 1950.

Figure \ref{BetaExplosurePlot} suggests some correlation between CMH and ``Betting against Beta'' (BAB), see \cite{Frazzini}. We indeed find that the CMH and BAB portfolios are only +10 \% correlated, 
such that the residue of the $\beta$-, Low-Vol- and BAB-neutralized CMH is still $4.7$.  Its weekly correlation with the residue of the $\beta$-, Low-Vol- and BAB-neutralized SMB is +71\%.

\begin{figure}
\centering
\includegraphics[width=10cm]{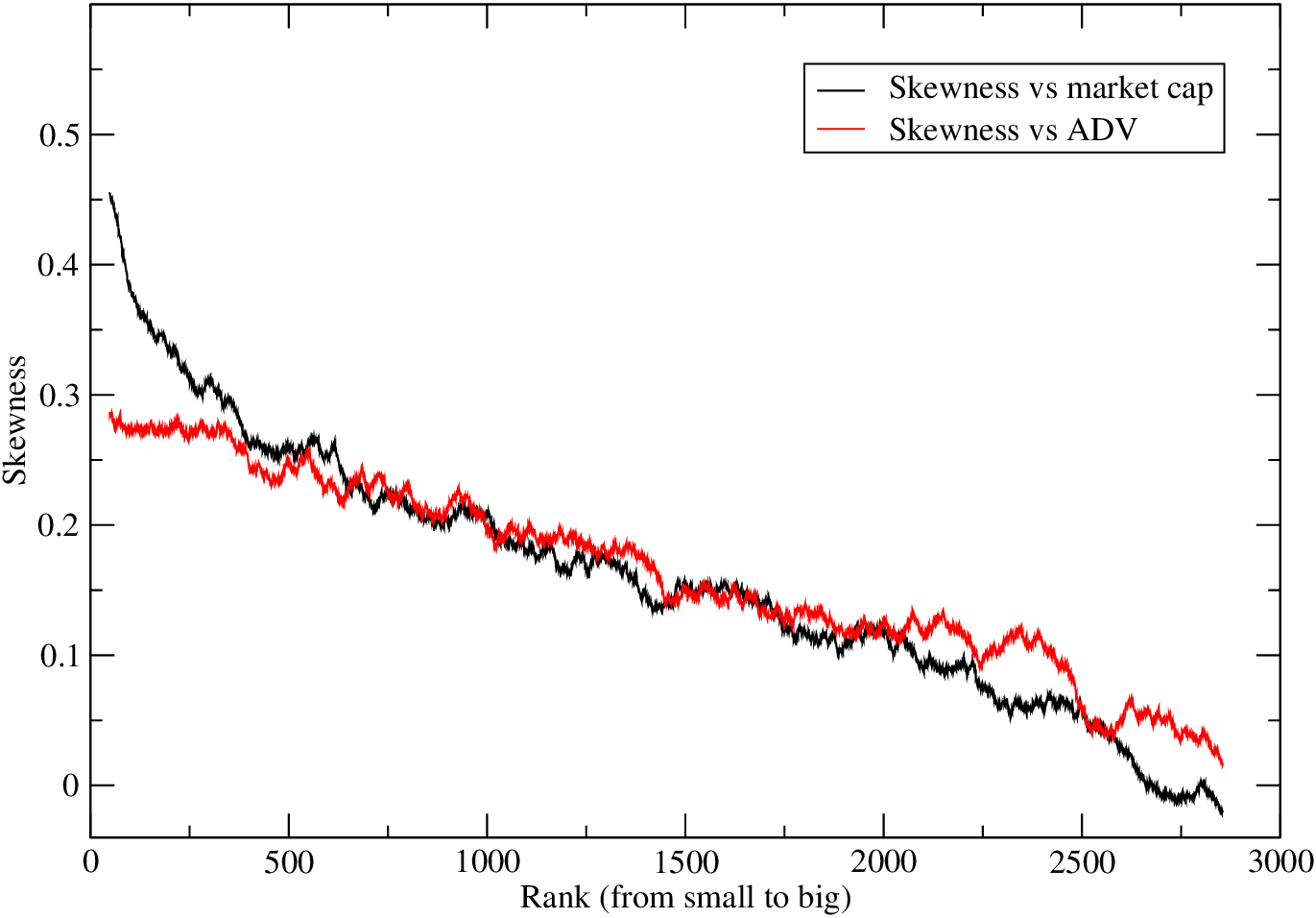}
\caption{Average single-name (daily) skewness vs rank (market cap or ADV). Small cap (ADV) stocks have a positive skewness on average.}
\label{skew_vs_rank}
\end{figure}

Finally, is CMH reducible to a liquidity risk premium, as measured by Amihud's LIQ indicator? The t-stat of a $\beta$-neutralised worldwide LIQ portfolio is $1.1$, only marginally larger than SMB ($0.9$). This is mostly because of the  presence of volatility in the denominator of LIQ: this means that portfolios that are short small LIQ and long large LIQ have a mechanical negative exposure to the Low-Vol effect. This results in a -55 \% correlation between LIQ and Low-Vol portfolios. Once Low-Vol-neutralised, LIQ portfolios are essentially equivalent to our CMH implementation. It would be interesting to extend this analysis to alternative measures of liquidity, see e.g. \cite{Sadka2008}.

\section{Is the Size Effect a Risk Premium?}

\begin{figure}
\centering
\includegraphics[width=10cm]{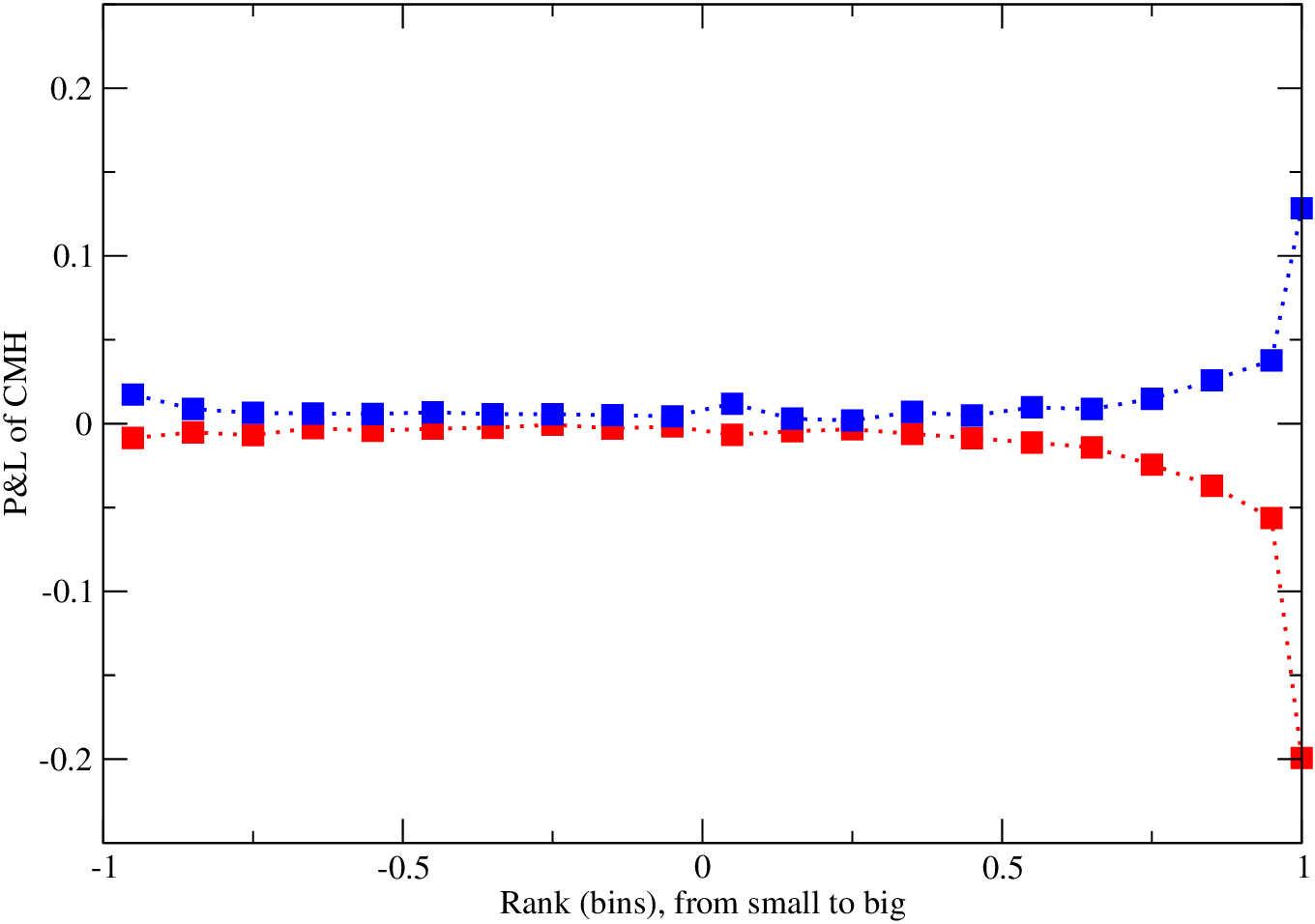}
\caption{Extreme portfolio returns decomposed on ADV bins. Here we consider the best 100 and the worst 100 daily PnLs out of a 20 years back-test on a US pool of stocks. For each day, the PnL of any decile is computed then averaged over the 100 days. Clearly, the largest contributions come from the largest decile of stocks for the most positive \emph{and} for the most negative days.
}
\label{CrashAnalysis}
\end{figure}

It is tempting to associate the size effect to a risk premium of sorts. Intuitively, it would make sense that investing in small cap stocks is more risky than in well established, large cap stocks. Correspondingly, small cap stocks should be less attractive and therefore undervalued. However, several empirical observations go against this naive picture. 
\begin{itemize}
\item First, as is now well known, volatility in itself does not entail risk premia, as the Low-Vol anomaly illustrates \cite{LowVol1,LowVol2,LowVol3} (and refs. therein). 
\item Second, the skewness of SMB portfolios is only weakly negative, and those of CMH virtually \emph{unskewed}, at odds with what one should expect for risk premium strategies (see the extended discussion in \cite{RiskPremia}). \item Third, the skewness of the returns of small caps/small ADV stocks is actually \emph{positive}, and decreases to zero as the market cap/ADV increases, see Figure \ref{skew_vs_rank}. 
\item Fourth, we find that extreme returns of the SMB or CMH portfolios are dominated by the largest caps/ADV -- see Figure \ref{CrashAnalysis}, indicating that the ``risk'' of such portfolios is not at all due to the exposure to small stocks, but rather to the supposedly safer large stocks. 
\item Fifth, as mentioned in the previous section, simple measures of liquidity risk have less explanatory power than size indicators themselves.
\end{itemize}

So, in what sense -- if any -- should the SMB/CMH effect be thought of as a risk reward? 

One possibility is that the size effect is not a risk premium but a behavioural anomaly. This could work as follows. Investors believe that small cap/ADV stocks have a attractive growth potential (reflected in the positive skewness of these stocks). They systematically buy these lottery tickets over several months, which would explain why Mutual Fund positions have a strong exposure to SMB, and why these stocks overperform for a while due to the sustained buy pressure, before eventually becoming themselves large-cap stocks. This explanation is however exactly the opposite of the reason why High Vol, glittering stocks should underperform.

\begin{figure}
\centering
\includegraphics[width=10cm]{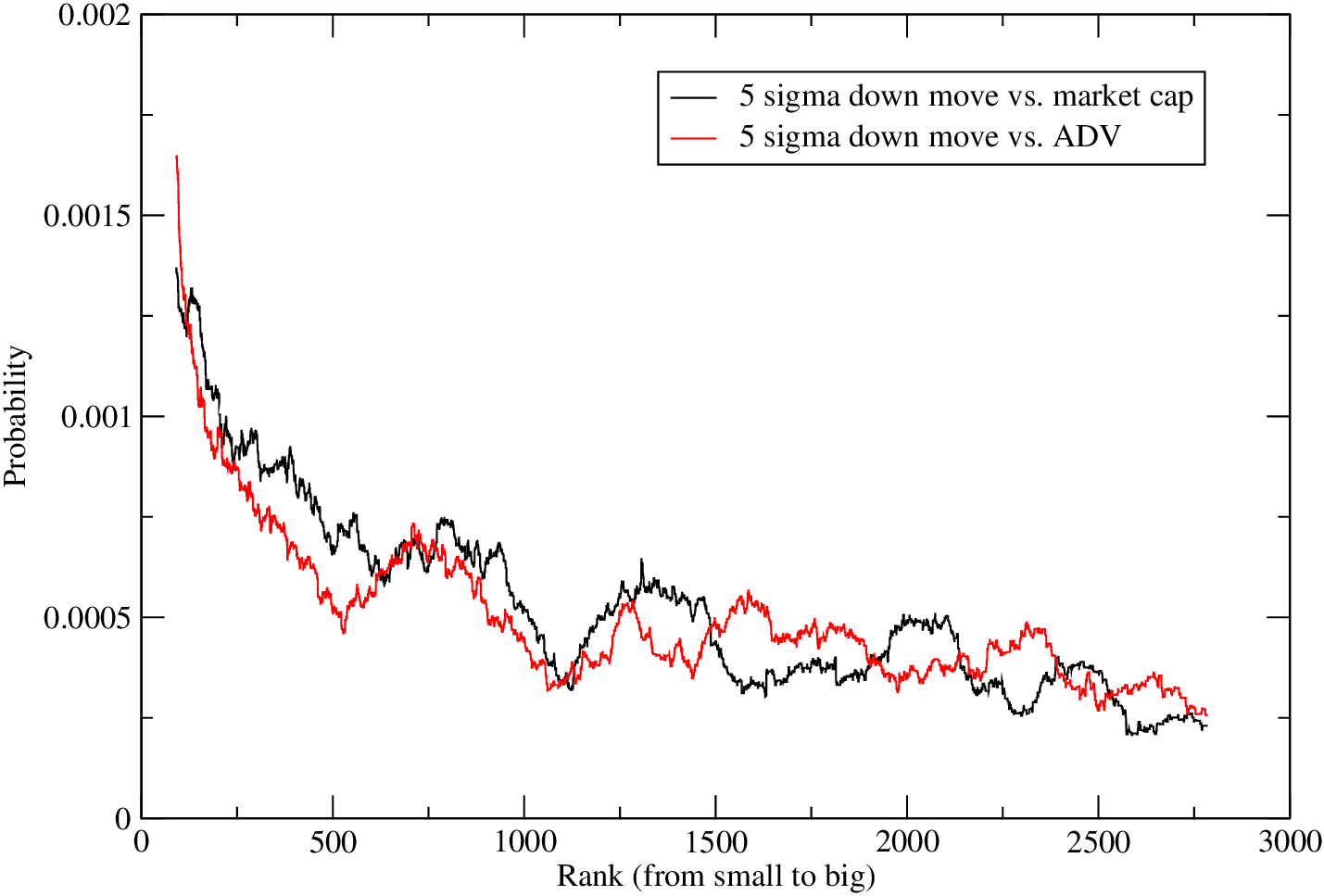}
\caption{Probability of a 5-sigma weekly draw-down for a stock vs rank (market cap or ADV). Although small stocks have a positive skewness, the probability of a 5-sigma draw-down is larger for small caps than for large caps.}
\label{down_move}
\end{figure}

Another story, more in line with the traditional risk premium lore, is that investors are in fact sensitive to higher moments of the return distribution, or to its extreme left-tail. The idea here is that even if the skewness is positive, the possible occurrence of large negative moves (leading to large losses) has a deterrent effect. We have studied two indicators capturing the excess probability of large moves. One is a (low-moment) kurtosis measure, which indeed shows a systematic downward trend with market capitalisation or ADV (data not shown). The second, perhaps more intuitive indicator is the probability of a 5-$\sigma$ weekly drawdown (where $\sigma$ is the weekly volatility, computed in a causal way). This probability is shown in Figure \ref{down_move}; it is seen to decline sharply as a function of both market cap and ADV, from around $2 \times 10^{-3}$ for small ADVs to $5 \times 10^{-4}$ for large ADVs. These negative tail events, even when less likely than their drawup counterparts, are perhaps where the ``risk'' of small stocks is hidden. This possibility is quite interesting from a theoretical point of view. It implies that a Taylor expansion of the utility function, as is often done to capture investors' preference, is unwarranted, since extreme tail events may dominate the average utility of an investment. In particular, such a Taylor expansion, truncated at third order, is used to argue why negative skewness assets are undervalued, and why positive skewness assets (lottery tickets) are overvalued. This would lead to the wrong conclusion in the case of small cap stocks, for which the skewness is positive. As a numerical illustration, consider an asset with negative returns uniformly distributed between $-100 \%$ and $0 \%$, and positive returns uniformly distributed between $0 \%$ and $300 \%$. The probabilities are chosen such that the mean return is zero. The rms is the found to be $100 \%$, and the skewness is \emph{positive}, equal to $3/2$. However, the average of a logarithmic utility function $U=\ln (1+r)$, where $r$ is the return, is negative, because large negative events are strongly penalized.

However, the above ``safety first'' interpretation is itself fraught with difficulties: according to the common lore, the individual risk of drawdowns being easily diversifiable (at least for professional investors), and should not be rewarded. Indeed, negative tail correlations are significantly larger among \emph{large cap} stocks than among small cap stocks. At the portfolio level, as we have emphasized, the largest contributors to extreme risks are also the largest cap/ADV stocks (see Figure \ref{CrashAnalysis}). The relevance of the probability of large drawdowns for explaining the existence of a size effect remains at best hypothetical.       

\section{Conclusion}

Our main conclusion is at odds with the idea that the size premium is a myth. On the contrary, we find that when measured in terms of dollar-turnover (ADV) rather than market capitalisation, and once $\beta$-neutralised and Low-Vol neutralised, the size effect is alive and well. With a long term t-stat of $5.1$, it is certainly not less significant than other well known factors such as Value or Quality. As compared to market-cap based SMB, ADV portfolios are much less anti-correlated to the Low-Vol anomaly. From a portfolio manager perspective, this should favour the CMH implementation over SMB, as the former is by construction less sensitive to volatility and $\beta$ biases and thus better suited for portfolio allocation.

The economic justification for the existence of the size effect is however only vaporous. As we have seen, the skewness of the returns of size-based portfolios are only weakly negative or insignificant. The extreme risk of these portfolios is dominated by the large caps, and the small caps have a positive (rather than negative) skewness. The only argument that favours a risk premium interpretation at the individual stock level is that the extreme drawdowns are more frequent for small cap/ADV stocks, even after accounting for volatility. Another, behavioural explanation could be a positive skewness (lottery ticket) mechanism, coupled with the long-term price impact of over-optimistic buyers -- but this explanation collides (at least naively) with the Low-Vol anomaly. More empirical work is clearly needed to decipher the size effect anomaly. 

\vskip 0.5cm
{\it Acknowledgments}. We thank Fabrizio Altarelli, Alexios Beveratos, Daniel Giamouridis and Laurent Laloux for many helpful suggestions.

\newpage
\appendix
\bibliographystyle{apalike}
\bibliography{SMB_biblio}

\begin{thebibliography}{}

\bibitem[Amihud, 2002]{Amihud_02}
Amihud, Y. (2002).
\newblock Illiquidity and stock returns: cross-section and time-series effects.
\newblock {\em Journal of Financial Markets}, 5(1):31--56.

\bibitem[Ang et~al., 2009]{LowVol1}
Ang, A., Hodrick, R.~J., Xing, Y., and Zhang, X. (2009).
\newblock High idiosyncratic volatility and low returns: International and
  further us evidence.
\newblock {\em Journal of Financial Economics}, 91:1--23.

\bibitem[Asness et~al., 2015]{Asness2015}
Asness, C., Frazzini, A., Israel, R., Moskowitz, T., and Pedersen, L. (2015).
\newblock Size matters, if you control your junk.
\newblock
  \url{https://www.aqr.com/library/working-papers/size-matters-if-you-control-your-junk}.

\bibitem[Baker et~al., 2011]{LowVol2}
Baker, M., Bradley, B., and Wurgler, J. (2011).
\newblock Benchmarks as limits to arbitrage: understanding the low-volatility
  anomaly.
\newblock {\em Financial Analysts Journal}, 67:40--54.

\bibitem[Banz, 1981]{Banz1981}
Banz, R.~W. (1981).
\newblock The relationship between return and market value of common stocks.
\newblock {\em Journal of Financial Economics}, 9(1):3--18.

\bibitem[Beveratos et~al., 2016]{LowVol3}
Beveratos, A., Bouchaud, J.-P., Ciliberti, S., Laloux, L., Lemp\'eri\`ere, Y.,
  Potters, M., and Simon, G. (2016).
\newblock Deconstructing the low-vol anomaly.
\newblock \url{https://arxiv.org/pdf/1510.01679.pdf}.

\bibitem[Blume and Stambaugh, 1983]{Blume1983}
Blume, M. and Stambaugh, R. (1983).
\newblock Biases in computed returns: An application to the size related
  effect.
\newblock {\em Journal of Financial Economics}, 12(3):387--404.

\bibitem[Brennan and Subrahmanyam, 1996]{BrennanSubra1996}
Brennan, M. and Subrahmanyam, A. (1996).
\newblock Market microstructure and asset pricing: On the compensation for
  illiquidity in stock returns.
\newblock {\em Journal of Financial Economics}, 41(3):441--464.

\bibitem[Chan and Chen, 1988]{ChanChen1988}
Chan, K. and Chen, N. (1988).
\newblock An unconditional asset-pricing test and the role of firm size as an
  instrumental variable for risk.
\newblock {\em Journal of Finance}, 43(2):309--325.

\bibitem[Chan et~al., 1985]{ChanHsieh1985}
Chan, K., Chen, N., and Hsieh, D. (1985).
\newblock An exploratory investigation of the firm size effect.
\newblock {\em Journal of Financial Economics}, 14(3):451--471.

\bibitem[Christie and Hertzel, 1981]{Christie1981}
Christie, A. and Hertzel, M. (1981).
\newblock Capital asset pricing "anomalies": Size and other correlations.

\bibitem[Fama and French, 1993]{FamaFrench1993}
Fama, E. and French, K. (1993).
\newblock Common risk factors in the returns on stocks and bonds.
\newblock {\em Journal of Financial Economics}, 33(1):3--56.

\bibitem[Frazzini and Pedersen, 2014]{Frazzini}
Frazzini, A. and Pedersen, L.~H. (2014).
\newblock Betting against beta.
\newblock {\em Journal of Financial Economics}, 111:1--25.

\bibitem[Huberman and Kandel, 1987]{Huberman1987}
Huberman, G. and Kandel, S. (1987).
\newblock Mean-variance spanning.
\newblock {\em Journal Finance}, 42(4):873--888.

\bibitem[Korajczyk and Sadka, 2008]{Sadka2008}
Korajczyk, R.~A. and Sadka, R. (2008).
\newblock Pricing the commonality across alternative measures of liquidity.
\newblock {\em Journal of Financial Economics}, 87(1):45--72.

\bibitem[Lemp\'eri\`ere et~al., 2017]{RiskPremia}
Lemp\'eri\`ere, Y., Deremble, C., Nguyen, T.-T., Seager, P., Potters, M., and
  Bouchaud, J.-P. (2017).
\newblock {Risk Premia: Asymmetric Tail Risks and Excess Return}.
\newblock {\em Quantitative Finance}, 17(1):1--14.

\bibitem[Reinganum, 1981]{Reinganum1981}
Reinganum, M.~R. (1981).
\newblock Misspecification of capital asset pricing: Empirical anomalies based
  on earnings' yields and market values.
\newblock {\em Journal of Financial Economics}, 9(1):19--46.

\bibitem[Roll, 1981]{Roll1981}
Roll, R. (1981).
\newblock A possible explanation of the small firm effect.
\newblock {\em Journal of Finance}, 36(4):879--88.

\bibitem[Schwert, 1983]{Schwert1983}
Schwert, G. (1983).
\newblock Size and stock returns, and other empirical regularities.
\newblock {\em Journal of Financial Economics}, 12(1):3--12.

\bibitem[van Dijk, 2011]{vanDijk2011}
van Dijk, M. (2011).
\newblock Is size dead? a review of the size effect in equity returns.
\newblock {\em Journal of Banking \& Finance}, 35(12):3263--3274.

\end{thebibliography}

\end{document}